\documentclass[aps,prl,preprint,groupedaddress]{revtex4-1}
\usepackage{color}
\draft 
\usepackage{graphicx}
\usepackage{caption}
\usepackage{float}
\usepackage{textcomp}

\begin{document}


\title{Enhanced thermoelectric properties of coaxial $\mathrm{Bi_2Te_3}$/$\mathrm{Sb_2Te_3}$ nanostructures studied by theoretical modeling}


\author{Qilin Gu}
\email[]{gump423@gmail.com}
\homepage[]{https://sites.google.com/site/qlgu423/}
\affiliation{Department of Electrical and Computer Engineering, The Ohio State University, Columbus, OH 43210, USA}


\date{December 12, 2010}

\begin{abstract}
Critical thermoelectric parameters including Seebeck coefficient, electrical conductivity, thermal conductivity and figure of merit \emph{ZT} of one-dimensional coaxial $\mathrm{Bi_2Te_3}$/$\mathrm{Sb_2Te_3}$ nanocomposite were modeled by following the single carrier pocket and sharp interface assumptions. A calculation scheme based on Landauer approach, instead of commonly used Boltzmann transport equation (BTE) with relaxation time approximation, was adopted to numerically obtain the transmission functions which can be used to evaluate thermoelectric properties. Considerable enhancement of \emph{ZT} was obtained through our modeling and numerical calculation, and the corresponding dependence of thermoelectric performance on structure parameters was studied. Finally, the efficiency at the maximum power condition for this 1-D system was also investigated.
\end{abstract}


\maketitle

As the demand for renewable energy resources increases dramatically over the last decade, the development of high-performance thermoelectric (TE) materials and related devices, which can be used to convert waste heat to electricity, has been attracting broader and greater interest.1 Enhancing the thermoelectric figure of merit \emph{ZT} in bulk materials remains difficult due to the inherent inter-couplings between three major physical parameters that determine \emph{ZT}, namely, Seebeck coefficient $\left|S\right|$, electrical conductivity $\sigma$ and thermal conductivity $\kappa$. Alternatively, low-dimensional nanostructured materials have been suggested to be promising candidates for improving thermoelectric efficiency based upon the proposed advantageous properties including the increased phonon scattering due to the high density of interfaces for reducing phonon thermal conductivity and the sharp features of electronic density of states for improving thermopower, which are absent in their bulk counterparts.\textcolor{blue}{$^{1}$} Thus, both theoretical and experimental studies of the thermoelectric properties of low-dimensional material systems such as superlattice and quantum dots have been pursued extensively.\textcolor{blue}{$^{1-4}$} 

Being one of the efforts to implement 1-D thermoelectric materials with high \emph{ZT} value using low-cost methods, core/shell nanocomposite structures have considerable potentials for tailoring electrical and thermal properties for practical device applications.\textcolor{blue}{$^{5-8}$} However, very few theoretical studies of these coaxial-type nanostructures have been reported. In the present study, we are herein trying to extend the general framework of thermoelectric modeling to 1-D core/shell structures to demonstrate optimal thermoelectric figures of merit by investigating the microscopic electronic and thermal transport properties of coaxial nanocomposites, which can potentially provide better understandings and reliable predictions of intrinsic properties of thermoelectric nanomaterials. $\mathrm{Bi_2Te_3}$/$\mathrm{Sb_2Te_3}$ heterostructure, a well-known material system which has already shown outstanding thermoelectric properties in its superlattice structure,\textcolor{blue}{$^{9}$} was chosen for this study. Fig. 1 (a) shows the schematic configuration of this coaxial nanocomposite whose total diameter is \emph{w}, consisting of a $\mathrm{Bi_2Te_3}$ core with diameter of \emph{d} surrounded by a $\mathrm{Sb_2Te_3}$ outer shell. The corresponding simplified band structure is also included to qualitatively show the quantum confinement effects that result in quantized sub-band levels as carrier pockets. It has been suggested that the inner and the outer diameters d and w in SiGe core/shell heterostructures had large impact on determining energy separations and the number of sub-band levels that are inside the core potential well respectively.\textcolor{blue}{$^{10}$} For simplicity, therefore, a diameter ratio (\emph{d/w}) of 1/2 has been used for all the following calculations, which can result in reasonably significant quantum confinement effects for improving thermoelectric properties. Additionally, single carrier pocket and sharp interface assumptions\textcolor{blue}{$^{10}$} have also been included in electronic transport calculations.

To model thermoelectric properties, TE physical parameters are commonly calculated by obtaining the integrals of transport distribution function through solving Boltzmann transport equation (BTE) in the relaxation time approximation.\textcolor{blue}{$^{11}$} In spite of the popularity of BTE-related TE modeling strategies, we instead adopted different calculation schemes based on Landauer approach which, as suggested by Jeong \emph{et al},\textcolor{blue}{$^{12}$} can provide accurate descriptions of TE parameters through simpler implementations. The Landauer expressions of basic electronic transport quantities electrical conductivity, Seebeck coefficienct, and the electronic thermal conductivity in linear response regime are given by\textcolor{blue}{$^{12-14}$} 
$$\sigma=\frac{2q^2}{h}I_0\frac{L}{A}\qquad [1/\mathrm{\Omega m}]$$  
$$S=\frac{k_B}{q}\frac{I_1}{I_0}\qquad [\mathrm{V/K}]$$
$$\kappa_e=\frac{2k_B^2T}{h}\left(I_2-\frac{I_1^2}{I_0}\right)\frac{L}{A}\qquad [\mathrm{W/Km}]$$
\noindent
where \emph{L} and \emph{A} are length and area of a diffusive conductor where electrical and thermal conductance \emph{G} and $K_e$ scale with length and can be expressed as electrical and thermal conductivity $\sigma$ and $\kappa$ respectively for convenience, and 
$$I_n=\int dE\left(\frac{E-E_F}{k_BT}\right)^n\left(-\frac{\partial{f}}{\partial{E}}\right)\left(\frac{\lambda_e(E)}{L}\right)M_e(E)$$
\noindent
with $\bar{T}_e$$(E)$=$(\lambda_e(E)/L)$$M_e(E)$ defined as the so-called transmission function,\textcolor{blue}{$^{15}$} $\lambda_e(E)$ being the mean-free-path for back-scattering of an electron with energy of E, $M_e(E)$ being the density of electron modes (DOEM) at energy E,\textcolor{blue}{$^{16}$} and $f$ being the Fermi function in the form of $f$=1/(1+$exp((E-E_F)$/$k_BT$)). It is worth noting that $\lambda_e(E)$/$L$ is an approximate form of $\lambda_e(E)$/$(\lambda_e(E)+L)$ in the diffusive limit. Furthermore, DOEM $M_e(E)$ in the Landauer formalism can be related to the electronic bandstructure and is essentially the product of the carrier velocity and the density of states (DOS). For 1-D nanowire case, assuming that a single parabolic subband is occupied, the density of electron modes can be evaluated by\textcolor{blue}{$^{17}$}
$$M_{1De}(E)=\Theta\left (E-\varepsilon_1\right)$$
\noindent
where $\Theta$ is the unit step function and $\varepsilon_1$ is the bottom of the first subband. On the other hand, the Landauer expression for phonon (lattice) thermal conductivity can be obtained through a similar treatment of phonons on a frequency by frequency basis:\textcolor{blue}{$^{18}$}
$$\kappa_p=\frac{\hbar^2}{2\pi k_BT^2}\int d\omega \omega^2 \frac{e^{\hbar\omega/k_BT}}{(e^{\hbar\omega/k_BT}-1)^2}\left(\frac{\lambda_p(\omega)}{T}\right)
M_p(\omega)\frac{L}{A}$$
\noindent
Here the photon transmission function can be written as $\bar{T}_p$($\omega$)=$(\lambda_p(\omega)/L)$$M_p$($\omega$), with $\lambda_p$($\omega$) being the mean-free-path for back-scattering of a phonon with a frequency $\omega$ and $M_p$($\omega$) being the density of phonon modes (DOPM) at a given frequency ¦Øwith a 1-D form of\textcolor{blue}{$^{18}$}
$$M_{1Dp}(\omega)=\Theta(\omega-\omega_i)(\omega_f-\omega)$$
\noindent
where $\omega_{i(f)}$ is the lower (upper) frequency limit of the first subband.

In order to verify the validity of our calculation schemes, a proof-of-concept comparison between our calculations and experimental observations are showed in Figure 1(b), where the calculated Seebeck coefficients (upper) and phonon thermal conductivities (lower) of \emph{n}-type $\mathrm{Bi_2Te_3}$ and $\mathrm{Sb_2Te_3}$ bulk materials are plotted along as a function of electrical conductivity and temperature respectively with corresponding experimental data from literature. The excellent agreements between our modeling and experiments indicate the feasibility of the following theoretical modeling.

Given the methods described above, critical thermoelectric transport coefficients of $\mathrm{Bi_2Te_3}$/$\mathrm{Sb_2Te_3}$ core/sell nanocomposites are modeled at 300 \emph{K} as a function of geometrical dimension - total wire diameter \emph{w} and relative Fermi level position with respect to the conduction (valence) band edge. Figure 2 (a) shows the dependence of figure of merit $ZT$ on the size of the coaxial structure and the Fermi energy. It is revealed that a considerably high peak $ZT$ value of $\scriptsize{\sim}$12 can be achieved at a relatively small outer diameter $w$$\scriptsize{\sim}$7.5 $nm$ and with a Fermi level located near valence band edge by \emph{p}-type doping. However, the desired high TE performance with $ZT$ values above 8 was calculated to take place only within a very narrow diameter ($w$) window between 5 $nm$ and 10 $nm$, and to drop significantly with increasing cross-section size. Further breakdown investigations on different component thermoelectric parameters that determine $ZT$ are also showed in Fig. 2(b)-(d) to explain the physical mechanism of calculated $ZT$ behaviors. Compared with its bulk counterpart, the calculated noticeably higher electrical conductivity $\sigma$ with the peak value of $\scriptsize{\sim}$1.6$\times$$10^7/\Omega m$ and enhanced Seebeck coefficient $\left|S\right|$ up to $\scriptsize{\sim}$600 $\mu V/K$ can be obtained (Fig. 2(b) and (c)), which is not surprising since the proposed sharp features in the electron density of states in 1-D systems due to the quantized electron motion perpendicular to the potential barrier are expected to produce increased electrical conductivity. The clear size dependences of both electrical conductivity and Seebeck coefficient, as shown in the contour plots, are believed to arise from the diameter modulated quantum confinement effects on the electronic band-structure of studied coaxial structure. Essentially, the separations between quantized energy levels and the consequent quantum effects reach their maxima at a small diameter value of $\scriptsize{\sim}$7.5 $nm$, and then decrease starkly when $w$ further increases, resulting in the strong diameter modulation of $\sigma$ and $\left|S\right|$. On the other hand, the Fermi energy dependence of shows that $\sigma$ and $\left|S\right|$ values peak near the band edges and decrease when the Fermi level $E_f$ further moves into the band. This $E_f$ dependence can be explained by realizing the opposite contributions to electrical conductivity and Seebeck coefficient from energy levels above and below Fermi level. The density of electron modes on both sides of $E_f$ will approach balanced when $E_f$ moves further into the bands, leading to a decreased net contribution to thermoelectric transport properties. A similar mechanism also applies in size and Fermi level dependence of thermal conductivity, especially the electronic contribution $\kappa_{el}$. The peak value of $\kappa_{el}$ matches the highest density of electron modes (DOEM), as predicted by the expression given above, while the maximum $\kappa_{el}$ value observed when $E_f$ locates in the band-gap can be explained by ambipolar heat conduction mechanism arising from the unequal energy flow carried by the balanced electron flow within each band.  

The temperature dependence of the calculated figure of merit and Seebeck coefficient of $\mathrm{Bi_2Te_3}$/$\mathrm{Sb_2Te_3}$ coaxial nanostructure has also been studied and plotted respectively along with that of bulk materials. The observed strong decreases of Seebeck coefficients (Fig. 3(b)) for bulk $\mathrm{Bi_2Te_3}$ and $\mathrm{Sb_2Te_3}$ materials when T$>$300 \emph{K} are believed to be a result of conductivity transition from doped to intrinsic nature at relatively high temperatures, while both the excess carriers and quantum confinement shift the peak value of nanostructure Seebeck coefficient to higher temperatures with respect to that of bulk materials by suppressing the intrinsic conductivity. Moreover, the differences in magnitude of $\left|S\right|$ and peak temperature for nanostructures with different wire cross sections and Fermi energy positions apparently result from the size and energy level dependence of quantum effect. Similarly, it's not difficult to understand the \emph{ZT} behavior in the presented temperature range, as shown in Fig. 3(a), by accounting for the temperature dependence of $\left|S\right|$, $\sigma$ and $\kappa$ To predict the practical performance of $\mathrm{Bi_2Te_3}$/$\mathrm{Sb_2Te_3}$ core/shell heterostructure, a generic thermoelectric device configuration consisting of a cold and a hot electron reservoir obeying a Fermi-Dirac contribution is considered, as shown in Fig. 4(a), where a symmetrical applied voltage difference is assumed for simplicity.\textcolor{blue}{$^{19}$} The two reservoirs are connected by the coaxial nanostructured TE device characterized by its transmission function, and the thermoelectric power will be generated when high-energy electrons move from the hot side to the empty states on the cold side. Besides, the temperatures for both sides were fixed to be $\mathrm{T_C}$=300 $K$ and $\mathrm{T_H}$=360 $K$ ($\Delta T/T=0.2$) for efficiency and power calculations in telluride-based 1D nanostructures. In addition to the thermoelectric efficiency ($\eta$), the efficiency at maximum power ($\eta_{maxP}$) under finite current condition has also been evaluated. The results in Fig. 4(b) show that the calculated $\eta$, $\eta_{maxP}$ and $P_{max}$ (scatters) from transmission function approach match the solid lines obtained using traditional $ZT$-based evaluations.\textcolor{blue}{$^{20}$} All the three quantities increase with increasing thermoelectric figure of merit, and the structure with the highest $ZT$ of $\scriptsize{\sim}$12 corresponds the highest achievable $\eta$ of 88\%*$\eta_C$ and $\eta_{maxP}$ of 50\%*$\eta_C$, approaching the Curzon-Ahlborn limit, the thermodynamically maximum efficiency under maximum condition.

In conclusion, thermoelectric coefficients and figure of merit of coaxial $\mathrm{Bi_2Te_3}$/$\mathrm{Sb_2Te_3}$ nanostructures with circular cross sections have been modeled by studying the corresponding transmission functions and related densities of modes for both electrons and phonons based on Landauer approach. Strong size, doping and temperature dependences of $\sigma$,$\left|S\right|$, $\kappa$ and $ZT$ suggest the crucial role of rational material design for reaching the theoretically achievable figure of merit up to 12 at 300 $K$. Additionally, the corresponding thermoelectric efficiency and power density of the studied core/shell nanostructure in practical TE devices have also been analyzed and $\eta_{maxP}$/$\eta_C$ is found to approach Curzon-Ahlborn efficiency limit of $\scriptsize{\sim}$51\% Carnot efficiency when the $ZT$ of $\mathrm{Bi_2Te_3}$/$\mathrm{Sb_2Te_3}$ nanocomposite increases to its peak value. 

\begin{acknowledgments}
The author wishes to thank Dr. Z. H. Qiao and Dr. Bin Wang for help discussions on numerical calculations  
\end{acknowledgments} 

\vfill
\newpage
\noindent
\underline{\textbf{Reference}}\\
$^1$L. D. Hicks and M. S. Dresselhaus, Phys. Rev. B \textbf{47}, 16631 (1993).\\
$^2$Y.-M. Lin, X. Sun, and M.S. Dresselhaus, Phys. Rev. B \textbf{62}, 4610 (2000).\\
$^3$Y.-M. Lin and M. S. Dresselhaus, Phys. Rev. B \textbf{68}, 075304 (2003).\\
$^4$A. I. Boukai, Y. Bunimovich, J. Tahir-Kheli, J.-K. Yu, W. A. Goddard III, and J. R. Heath, Nature \textbf{451}, 168 (2008).\\
$^5$K. F. Hsu, S. Loo, F. Guo, W. Chen, J. S. Dyck, C. Uher, T. Hogan, E. K. Polychroniadis, and M. G. Kanatzidis, Science \textbf{303}, 818 (2004)\\
$^6$J. R. Sootsman, R. J. Pcionek, H. Kong, C. Uher, and M. G. Kanatzidis, Chem. Mater. \textbf{18}, 4993 (2006).\\
$^7$B. Poudel, Q. Hao, Y. Ma, Y. Lan, A. Minnich, B. Yu, X. Yan, D. Wang, A. Muto, D. Vashaee, X. Chen, J. Liu, M. S. Dresselhaus, G. Chen, and Z. F. Ren, Science \textbf{320}, 634 (2008).\\
$^8$Y. Ma, Q. Hao, B. Poudel, Y. Lan, B. Yu, D. Wang, G. Chen, and Z. F. Ren, Nano Lett. \textbf{8}, 2580 (2008).\\
$^9$R. Venkatasubramanian, E. Siivola, T. Colpitts and B. O'Quinn, Nature \textbf{413}, 597 (2001).\\
$^{10}$M. Y. Tang, Master thesis, Massachusetts Institute of Technology, 2004.\\
$^{11}$G. D. Mahan and J. O. Sofo, Proc. Natl Acad. Sci. USA \textbf{93}, 7436-7439 (1996).\\
$^{12}$C. Jeong, R. Kim, M. Luisier, S. Datta, and M. Lundstrom, J. Appl. Phys. \textbf{107}, 023707 (2010).\\
$^{13}$U. Sivan and Y. Imry, Phys. Rev. B \textbf{33}, 551 (1986).\\
$^{14}$C. R. Proetto, Phys. Rev. B \textbf{44}, 9096 (1991).\\
$^{15}$S. Datta, \emph{Electronic Transport in Mesoscopic System} (Cambridge University Press, New York, 1997).\\
$^{16}$S. Datta, \emph{Quantum Transport : Atom to Transistor} (Cambridge University Press, New York, 2005).\\
$^{17}$R. Kim, S. Datta, and M. S. Lundstrom, J. Appl. Phys. \textbf{105}, 034506 (2009).\\
$^{18}$T. Markussen, A. Jauho and M. Brandbyge, Phys. Rev. B \textbf{79}, 035415 (2009).\\
$^{19}$N. Nakpathomkun, H. Q. Xu, and H. Linke, Phys. Rev. B \textbf{82}, 235428 (2010).\\
$^{20}$G. S. Nolas, J. Sharp, and H. J. Goldsmid, \emph{Thermoelectrics: Basic Principles and New Materials Developments} (Springer,2001).\\

\begin{figure}[H]
\centering
\includegraphics[width=1\textwidth]{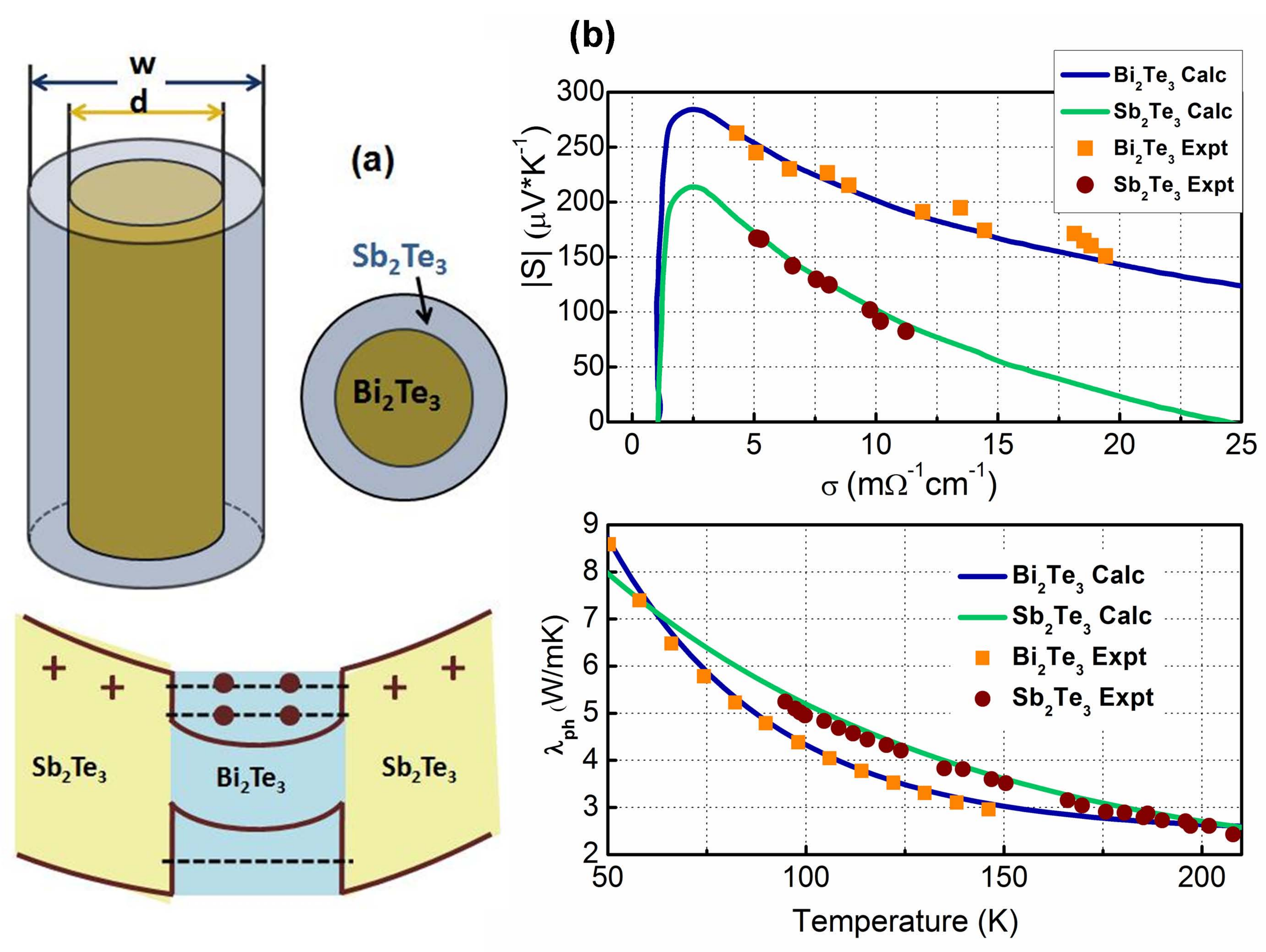}
\caption{\label{FIG.1} Schematic diagram of modeled $\mathrm{Bi_2Te_3}$/$\mathrm{Sb_2Te_3}$ 1-D system: prospective view, cross-section view and band diagram. (b) Comparison of Seebeck coefficients as a function of electrical conductivity (upper) and phonon thermal conductivity as a function of temperature (lower) for $\mathrm{Bi_2Te_3}$ and $\mathrm{Sb_2Te_3}$ respectively obtained from experiments and the calculation presented in this study.}
\end{figure}
\vfill
\newpage

\begin{figure}[H]
\centering
\includegraphics[width=0.9\textwidth]{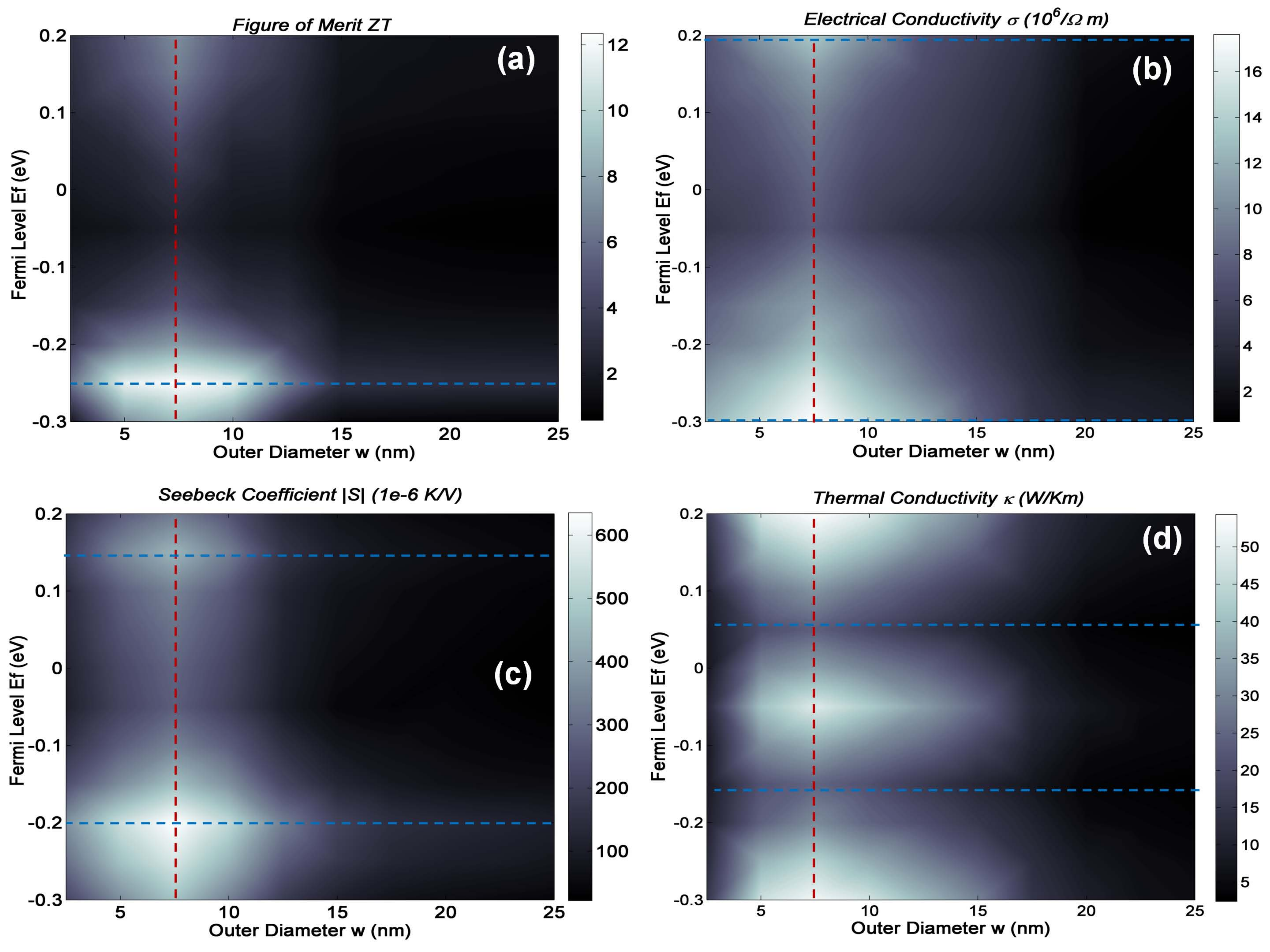}
\caption{\label{FIG.2} Thermoelectric parameters including figure of Merit \emph{ZT} (a), electrical conductivity $\sigma$ (b), Seebeck coefficient $\left|S\right|$ (c) and thermal conductivity $\kappa$ (d) are plotted as functions of the outer diameter of coaxial nanostructures and Fermi level energy. Dashed lines represent peak or valley values of corresponding parameters}
\end{figure}
\vfill
\newpage
\begin{figure}[H]
\centering
\includegraphics[width=1\textwidth]{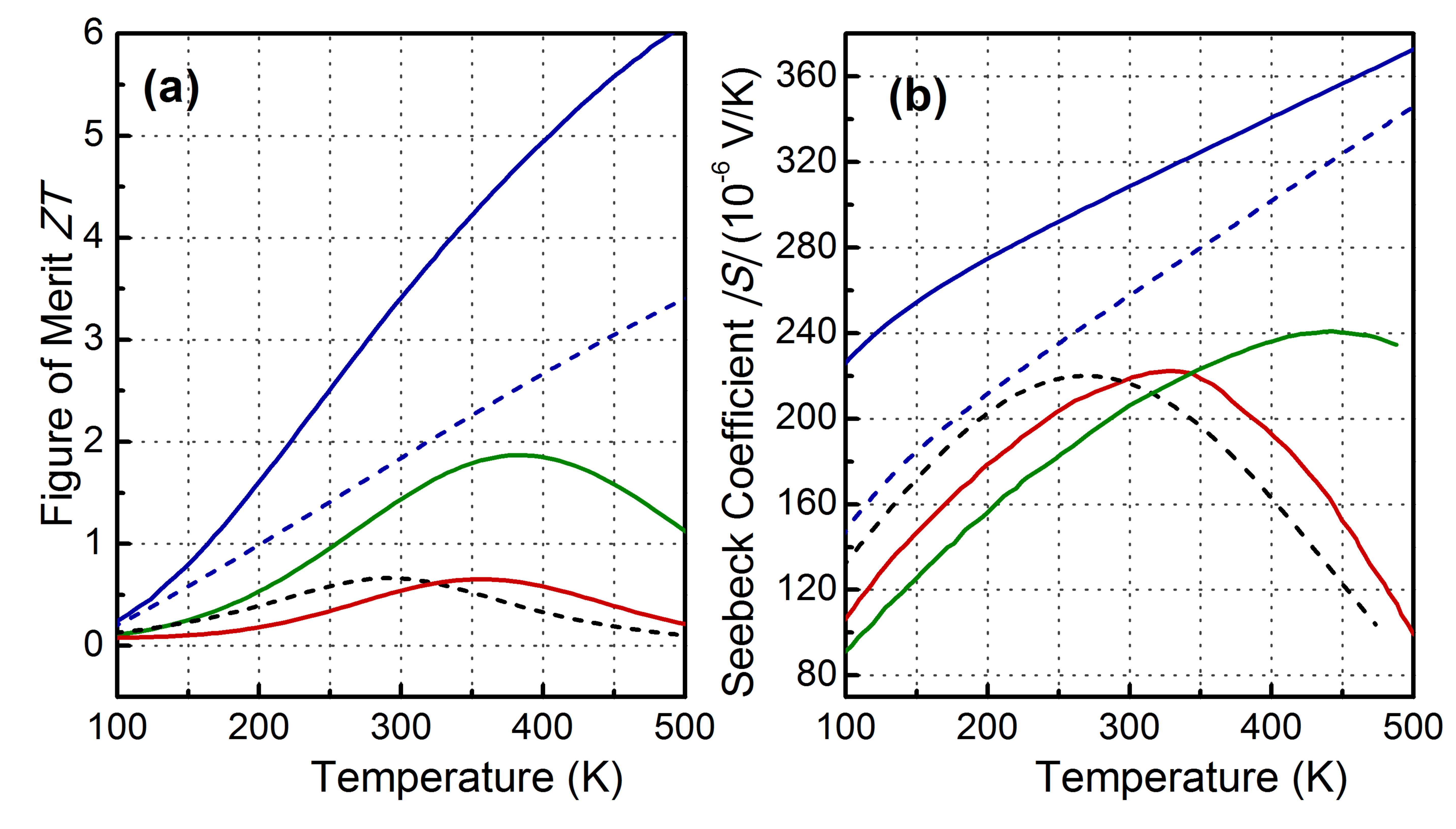}
\caption{\label{FIG.3} Temperature dependence of figure of merit \emph{ZT} (a) and Seebeck coefficient $\left|S\right|$ (b) for bulk $\mathrm{Bi_2Te_3}$ (red solid line), bulk $\mathrm{Sb_2Te_3}$ (black dashed line), $\mathrm{Bi_2Te_3}$/$\mathrm{Sb_2Te_3}$ core/shell with $w$=20 $nm$ (green solid line), $w$=7.5 $nm$ with $E_f$=-0.05 $eV$ (blue dashed line) and $w$=7.5 nm with $E_f$=-0.1 $eV$ (blue solid line).}
\end{figure}
\vfill
\newpage

\begin{figure}[H]
\centering
\includegraphics[width=1\textwidth]{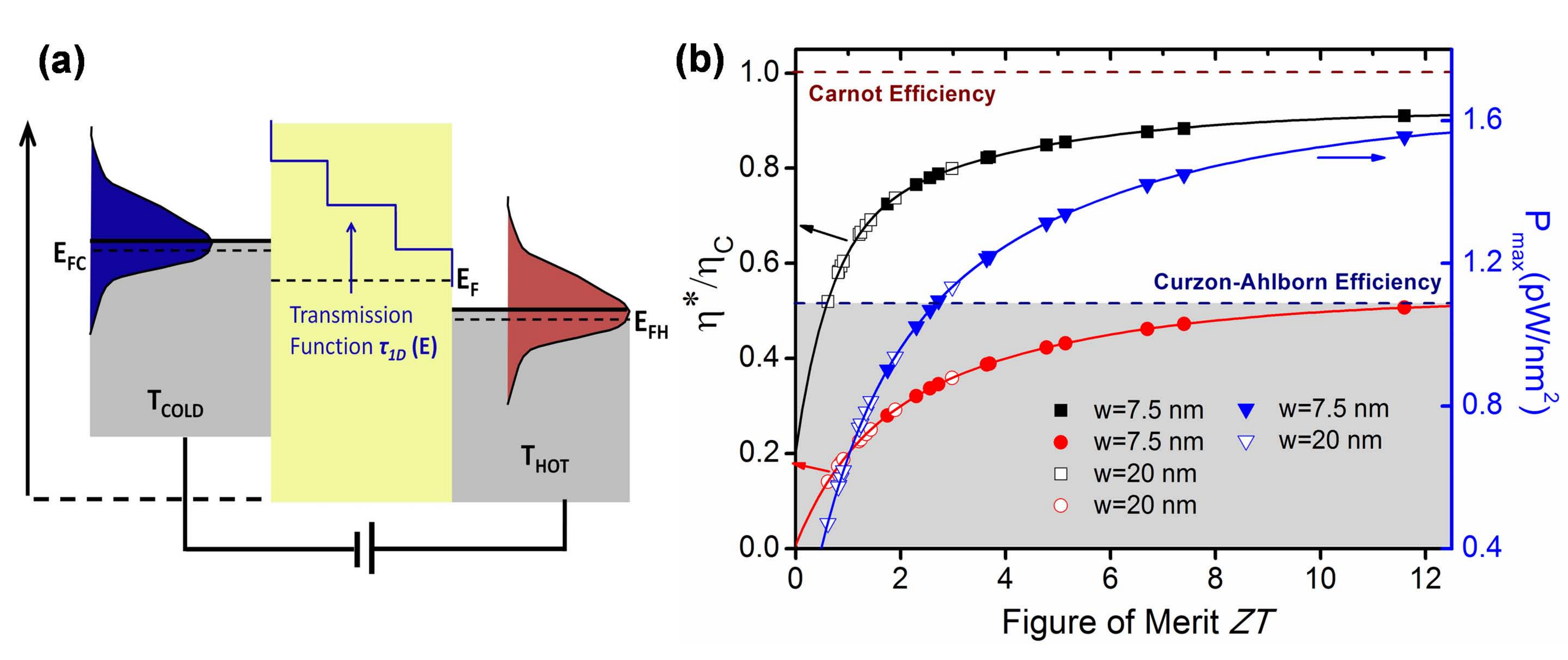}
\caption{\label{FIG.4} Schematic diagram set-up of a representative thermoelectric device which can be described by its transmission function $\bar{T}_{1D}$(E), as sketched for example, with contact leads acting as a cold and a hot electron reservoirs; (b) thermoelectric maximum efficiency (black square), the efficiency at maximum power (red circle), both normalized by Carnot efficiency (also labeled in the figure), and maximum power density are plotted as functions of figure of merit \emph{ZT}.}
\end{figure}

\subsection{}
\subsubsection{}

\bibliography{}

\end{document}